\documentclass[copyright,creativecommons]{eptcs}
\providecommand{\event}{THedu 2011}
\usepackage{breakurl}
\usepackage{graphicx}
\usepackage{isabelle,isabellesym}

\isadroptag{theory}
\isabellestyle{it}
\renewcommand{\isadigit}[1]{\isatext{#1}}

\newcommand{\bgtt}{\bgroup\isabellestyle{default}\isabellestyle{tt}\isastyle%
\renewcommand{\isacharverbatimopen}{\isamath{\langle\!\langle}}%
\renewcommand{\isacharverbatimclose}{\isamath{\rangle\!\rangle}}%
\renewcommand{\isadigit}[1]{##1}}
\newcommand{\entt}{\egroup}
\renewenvironment{isatagML}{\bgtt}{\entt}

\def\titlerunning{Isabelle/PIDE as Platform for Educational Tools}
\def\authorrunning{M. Wenzel \& B. Wolff}

\begin{document}

\title{Isabelle/PIDE as Platform for Educational Tools}
\author{Makarius Wenzel and Burkhart Wolff \\
  \institute{Universit\'e Paris-Sud 11, LRI, Orsay, France}}
\maketitle

\begin{abstract}
  The Isabelle/PIDE platform addresses the question whether proof
  assistants of the LCF family are suitable as technological basis for
  educational tools.  The traditionally strong logical foundations of
  systems like HOL, Coq, or Isabelle have so far been counter-balanced
  by somewhat inaccessible interaction via the TTY (or minor variations
  like the well-known Proof~General / Emacs interface).  Thus the
  fundamental question of math education tools with fully-formal
  background theories has often been answered negatively due to
  accidental weaknesses of existing proof engines.

  The idea of ``PIDE'' (which means ``Prover IDE'') is to integrate
  existing provers like Isabelle into a larger environment, that
  facilitates access by end-users and other tools.  We use Scala to
  expose the proof engine in ML to the JVM world, where many
  user-interfaces, editor frameworks, and educational tools already
  exist.  This shall ultimately lead to combined mathematical
  assistants, where the logical engine is in the background, without
  obstructing the view on applications of formal methods, formalized
  mathematics, and math education in particular.
\end{abstract}

\begin{isabellebody}%
\def\isabellecontext{Paper}%
\isadelimtheory
\endisadelimtheory
\isatagtheory
\isacommand{theory}\isamarkupfalse%
\ Paper\isanewline
\isakeyword{imports}\ Main\isanewline
\isakeyword{begin}%
\endisatagtheory
{\isafoldtheory}%
\isadelimtheory
\endisadelimtheory
\isamarkupsection{Introduction%
}
\isamarkuptrue%
\begin{isamarkuptext}%
\begin{quote}
  \small\em

  ``Isabelle's user interface is no advance over LCF's, which is
  widely condemned as `user-unfriendly': hard to use, bewildering to
  beginners. Hence the interest in proof editors, where a proof can be
  constructed and modified rule-by-rule using windows, mouse, and
  menus. But Edinburgh LCF was invented because real proofs require
  millions of inferences. Sophisticated tools -- rules, tactics and
  tacticals, the language ML, the logics themselves -- are hard to
  learn, yet they are essential. We may demand a mouse, but we need
  better education and training.''

  (L.C. Paulson, in ``Isabelle: The Next 700 Theorem Provers'', 1990)

  \end{quote}%
\end{isamarkuptext}%
\isamarkuptrue%
\isamarkupsubsection{The LCF prover family%
}
\isamarkuptrue%
\begin{isamarkuptext}%
Isabelle \cite{Wenzel-Paulson-Nipkow:2008} is an interactive
  theorem prover platform in the tradition of LCF
  \cite{Gordon-Milner-Wadsworth:1979}.  Other notable descendants of
  LCF are HOL and Coq, see also \cite{Wiedijk:2006}.  Even after
  several decades, current systems share the following main traits of
  the original LCF approach:

  \begin{enumerate}

  \item \textbf{Strong logical foundations.} Some well-understood
  logical basis is taken as starting point, and mathematical theories
  are explicitly constructed by reduction to first principles.  This
  follows the tradition of ``honest toil'' in the sense of Bertrand
  Russel: results are not just postulated as axioms, but derived from
  definitions as actual theorems.

  \item \textbf{Free programmability and extensibility.} Derived proof
  tools can be implemented on top of the logical core, while fully
  retaining its integrity.  This works by the strong type-safety
  properties of the ML implementation platform of the prover.
  
  \item \textbf{Primitive read-eval-print loop.} User interaction
  works by issuing individual commands, which the prover interprets on
  the spot and prints results accordingly.  Prover commands update an
  implicit state, which most newer provers allow to \emph{undo} in a
  linear fashion, to support the well-known proof scripting mode of
  Proof~General \cite{Aspinall:2000}.

  \end{enumerate}

  On the one hand, this general architecture proved quite successful
  in building reasonably large libraries of formal theories (such as
  the Archive of Formal Proof\footnote{\url{http://afp.sf.net}} for
  Isabelle).  On the other hand, integrating an LCF-style prover into
  into a combined mathematical assistant, especially one intended for
  math education, poses some challenges:

  \begin{description}

  \item[Interaction.] How can casual users interact with the prover,
  without getting exposed to the full details of logic implemented on
  the computer?  How can specific interaction scenarios that are
  relevant for computer-assisted math education be supported, hiding
  the fact that there is a fully-featured prover engine at the bottom?

  \item[Integration.] How can other systems connect to the prover
  engine?  How can we overcome the traditional plumbing of the
  read-eval-print loop via pipes, typically with synchronous /
  sequential protocols.  How can we proceed to the next generation of
  integrated mathematical assistants, with sophisticated front-ends
  and back-ends, using asynchronous / parallel evaluation?

  \end{description}

  In the past few years, Isabelle has acquired more and more support
  for \emph{Prover IDE} concepts, to address the above issues
  systematically.  This facilitates future implementations of
  mathematical assistants and applications to math education.  In
  particular, the read-eval-print loop is replaced by a \emph{document
  model} for direct editing with asynchronous interaction and parallel
  checking.  This is a continuation of earlier work on parallel proof
  checking \cite{Wenzel:2009,Matthews-Wenzel:2010} and modern prover
  interfaces \cite{Wenzel:2010}.  Some aspects of formal
  document-content have already been covered in
  \cite{Wenzel:2011:CICM}.  In the present paper, we discuss the
  overall PIDE framework and its technological side-conditions as
  integrative platform for educational tools that happen to care about
  logical foundations.%
\end{isamarkuptext}%
\isamarkuptrue%
\isamarkupsubsection{Formal logic for education%
}
\isamarkuptrue%
\begin{isamarkuptext}%
Formality in educational tools cannot be taken for granted.
  Major teaching tools for dynamic geometry like
  Geogebra\footnote{\url{http://www.geogebra.org}} have managed to
  ignore formal background theory for many years.  This raises the
  general question if the informality of contemporary math education
  tools is substantial or accidental.

  \medskip \emph{Substantial} lack of formal foundations would mean
  that logical principles are considered irrelevant or unwanted for
  math education.  Even the principle of mathematical proof appears to
  be challenged occasionally, and removed from many curricula of
  high-school mathematics.  This could be the starting point of
  philosophical and political discussions of the meaning of profound
  mathematical understanding for our technological society, but that
  is beyond the scope of the present paper.

  \emph{Accidental} lack of formal foundations can be explained easily
  by technical side-conditions.  Inaccessible and arcane theorem
  provers are unlikely to be considered as a platform for educational
  tools.  Classic Isabelle has not been any better than HOL or Coq in
  this respect.  In the past, there have been attempts to make
  LCF-style provers easier to access for their own right nonetheless,
  but the resulting culture of crude user-interfaces for theorem
  provers is still confined by the interaction model of Proof~General
  \cite{Aspinall:2000}.  This proved successful in its time, because
  it fits tightly with the TTY-based read-eval-print loop of existing
  systems, but it was difficult to continue from there.  The PGIP
  initiative \cite{Aspinall-et-al:2007}, which was essentially based on
  the same old command line model, could not repeat the acceptance of
  classic Proof General~/ Emacs, even though Eclipse was propagated as
  new front-end technology.

  \medskip Our PIDE approach is meant to change the rules of the game
  again, to expose interactive theorem proving to a broader audience,
  especially for education in mathematics, and other disciplines that
  depend on mathematical and logical foundations (e.g.\ formal-methods
  or systems engineering).  By overcoming the technical restrictions
  of interactive provers wrt.\ interaction and integration into
  educational tools, it should eventually become possible to
  investigate formal foundations of math education again.

  \medskip In the present paper, we outline the current status of the
  Isabelle/PIDE platform with its approach to \emph{domain-specific
  formal languages} for embeddings into the logical environment that
  are accessible to IDE front-ends.  This includes some architecture
  for structured specifications and proofs (Isabelle/Isar), parallel
  symbolic computation (Isabelle/ML), and higher-order object-oriented
  system integration (Isabelle/Scala).  Moreover, the current release
  of Isabelle2011-1 (October
  2011)\footnote{\url{http://isabelle.in.tum.de}} already includes
  Isabelle/jEdit as concrete Prover IDE implementation within the
  generic framework.  We shall also sketch further potential
  application scenarios, especially for educational purposes.%
\end{isamarkuptext}%
\isamarkuptrue%
\isamarkupsection{Isabelle/Isar proof documents%
}
\isamarkuptrue%
\begin{isamarkuptext}%
One of the specific strengths of the Isabelle platform is the
  Isar proof language \cite{Wenzel:1999} that allows to express formal
  reasoning in a way that is both human-readable and
  machine-checkable.  Unlike other systems in this category (notably
  Mizar \cite{Wiedijk:2006}), the Isar proof language is merely an
  application of a more general framework for structured logical
  environments: the notion of \emph{Isar proof context} provides
  flexible means to operate within the scope of locally fixed
  parameters and assumptions.  Over the years the general framework
  has already been re-used for structured specifications and module
  systems in Isabelle (such as locales and type-classes
  \cite{Haftmann-Wenzel:2006:classes,Haftmann-Wenzel:2009}).

  Beyond the default Isar proof language, it is possible to implement
  domain-specific languages for structured reasoning.  To illustrate
  some possibilities, we show outlines for calculational reasoning and
  induction in Isabelle/Isar.  The subsequent examples use the recent
  \textbf{notepad} element that allows to sketch formal reasoning
  independently of any pending goals.\medskip%
\end{isamarkuptext}%
\isamarkuptrue%
\begin{center}\begin{minipage}{0.35\textwidth}
\isacommand{notepad}\isamarkupfalse%
\isanewline
\isakeyword{begin}\isanewline
\isadelimproof
\ \ %
\endisadelimproof
\isatagproof
\isacommand{have}\isamarkupfalse%
\ {\isaliteral{22}{\isachardoublequoteopen}}a\ {\isaliteral{3D}{\isacharequal}}\ b{\isaliteral{22}{\isachardoublequoteclose}}\ \isacommand{sorry}\isamarkupfalse%
\isanewline
\ \ \isacommand{also}\isamarkupfalse%
\isanewline
\ \ \isacommand{have}\isamarkupfalse%
\ {\isaliteral{22}{\isachardoublequoteopen}}{\isaliteral{2E}{\isachardot}}{\isaliteral{2E}{\isachardot}}{\isaliteral{2E}{\isachardot}}\ {\isaliteral{3D}{\isacharequal}}\ c{\isaliteral{22}{\isachardoublequoteclose}}\ \isacommand{sorry}\isamarkupfalse%
\isanewline
\ \ \isacommand{also}\isamarkupfalse%
\isanewline
\ \ \isacommand{have}\isamarkupfalse%
\ {\isaliteral{22}{\isachardoublequoteopen}}{\isaliteral{2E}{\isachardot}}{\isaliteral{2E}{\isachardot}}{\isaliteral{2E}{\isachardot}}\ {\isaliteral{3D}{\isacharequal}}\ d{\isaliteral{22}{\isachardoublequoteclose}}\ \isacommand{sorry}\isamarkupfalse%
\isanewline
\ \ \isacommand{finally}\isamarkupfalse%
\isanewline
\ \ \isacommand{have}\isamarkupfalse%
\ {\isaliteral{22}{\isachardoublequoteopen}}a\ {\isaliteral{3D}{\isacharequal}}\ d{\isaliteral{22}{\isachardoublequoteclose}}\ \isacommand{{\isaliteral{2E}{\isachardot}}}\isamarkupfalse%
\endisatagproof
{\isafoldproof}%
\isadelimproof
\isanewline
\endisadelimproof
\isacommand{end}\isamarkupfalse%
\end{minipage}\begin{minipage}{0.4\textwidth}
\isacommand{notepad}\isamarkupfalse%
\isanewline
\isakeyword{begin}\isanewline
\isadelimproof
\ \ %
\endisadelimproof
\isatagproof
\isacommand{fix}\isamarkupfalse%
\ n\ {\isaliteral{3A}{\isacharcolon}}{\isaliteral{3A}{\isacharcolon}}\ nat\ \isacommand{have}\isamarkupfalse%
\ {\isaliteral{22}{\isachardoublequoteopen}}P\ n{\isaliteral{22}{\isachardoublequoteclose}}\isanewline
\ \ \isacommand{proof}\isamarkupfalse%
\ {\isaliteral{28}{\isacharparenleft}}induct\ n{\isaliteral{29}{\isacharparenright}}\isanewline
\ \ \ \ \isacommand{case}\isamarkupfalse%
\ {\isadigit{0}}\ \isacommand{show}\isamarkupfalse%
\ {\isaliteral{22}{\isachardoublequoteopen}}P\ {\isadigit{0}}{\isaliteral{22}{\isachardoublequoteclose}}\ \isacommand{sorry}\isamarkupfalse%
\isanewline
\ \ \isacommand{next}\isamarkupfalse%
\isanewline
\ \ \ \ \isacommand{case}\isamarkupfalse%
\ {\isaliteral{28}{\isacharparenleft}}Suc\ n{\isaliteral{29}{\isacharparenright}}\isanewline
\ \ \ \ \isacommand{from}\isamarkupfalse%
\ {\isaliteral{60}{\isacharbackquoteopen}}P\ n{\isaliteral{60}{\isacharbackquoteclose}}\ \isacommand{show}\isamarkupfalse%
\ {\isaliteral{22}{\isachardoublequoteopen}}P\ {\isaliteral{28}{\isacharparenleft}}Suc\ n{\isaliteral{29}{\isacharparenright}}{\isaliteral{22}{\isachardoublequoteclose}}\ \isacommand{sorry}\isamarkupfalse%
\isanewline
\ \ \isacommand{qed}\isamarkupfalse%
\endisatagproof
{\isafoldproof}%
\isadelimproof
\isanewline
\endisadelimproof
\isacommand{end}\isamarkupfalse%
\end{minipage}\end{center}
\begin{isamarkuptext}%
\bigskip \noindent Here the language elements \textbf{also}
  and \textbf{finally} for calculational sequences, and the specific
  \emph{induct} proof method with its symbolic cases are defined in
  user-space, although they happen to be part of the standard library
  of Isabelle.  By understanding Isabelle/Isar as generic framework
  for domain-specific formal languages, rather than a particular proof
  language, tool builders can start to implement their own
  sub-languages.  See also the main Isabelle/Isar reference manuals
  \cite{isabelle-isar-ref,isabelle-implementation} for further
  details.

  \medskip The idea of integrating a variety of formal languages into
  Isar proof documents has been there from the beginning
  \cite{Wenzel:1999}, and continues the ``generic framework''
  tradition of early Isabelle \cite{paulson700}.  In the past this was
  mostly limited to command-line interaction and batch-processing of
  plain text sources, with some off-line presentation in PDF-\LaTeX.
  In Isabelle/PIDE the concept of formal proof document is extended
  for online interaction, with continuous proof checking and
  visualization of formal content in the editor (semantic
  highlighting, tooltips, popups, hyperlinks etc.).

  In order to make this work systematically, the traditional prover
  engine is connected to the JVM, using the Isabelle/Scala integration
  layer that is already part of recent Isabelle versions.%
\end{isamarkuptext}%
\isamarkuptrue%
\isamarkupsection{Isabelle/ML versus Isabelle/Scala%
}
\isamarkuptrue%
\begin{isamarkuptext}%
Isabelle/ML is both the implementation and extension language
  of Isabelle/Isar.  ML is embedded into the formal Isar context, such
  that user code can refer to formal entities in the text, to achieve
  some static type-checking of logical syntax
  \cite{Wenzel-Chaieb:2007b}. It is also possible to make ML tools
  depend on logical parameters and assumptions, and apply them later
  in a different context with concrete terms and theorems
  \cite{Chaieb-Wenzel:2007}.

  The relation of Isabelle/ML versus the Isabelle logical framework is
  best understood as an elaboration of the original LCF approach
  \cite{Gordon-Milner-Wadsworth:1979}, where ML was introduced as
  meta-language to manipulate logical entities with full access to
  formal syntax.

  \medskip Isabelle/Scala is a new layer around the Isabelle prover
  that was introduced to facilitate system interaction and integration
  as discussed before.  The idea is to wrap the traditional ML prover
  process into some library for Scala/JVM \cite{Scala:2004}.  Thus the
  prover can be accessed via some statically-typed Scala API, while
  the connection between the two different processes (ML back-end
  versus JVM front-end) is hidden in some \emph{internal protocol}.
  The implementation leverages the existing concurrent and parallel
  infrastructure of both ML \cite{Matthews-Wenzel:2010} and Scala
  \cite{Haller-Odersky:2006}.

  Our principle of \emph{public API} and \emph{private protocol} turns
  out as important to achieve high-performance prover communication
  and flexibility for future refinements of the interaction model
  (precise evaluation strategies for document sub-structure etc.).

  \begin{center}
  \includegraphics[scale=0.7]{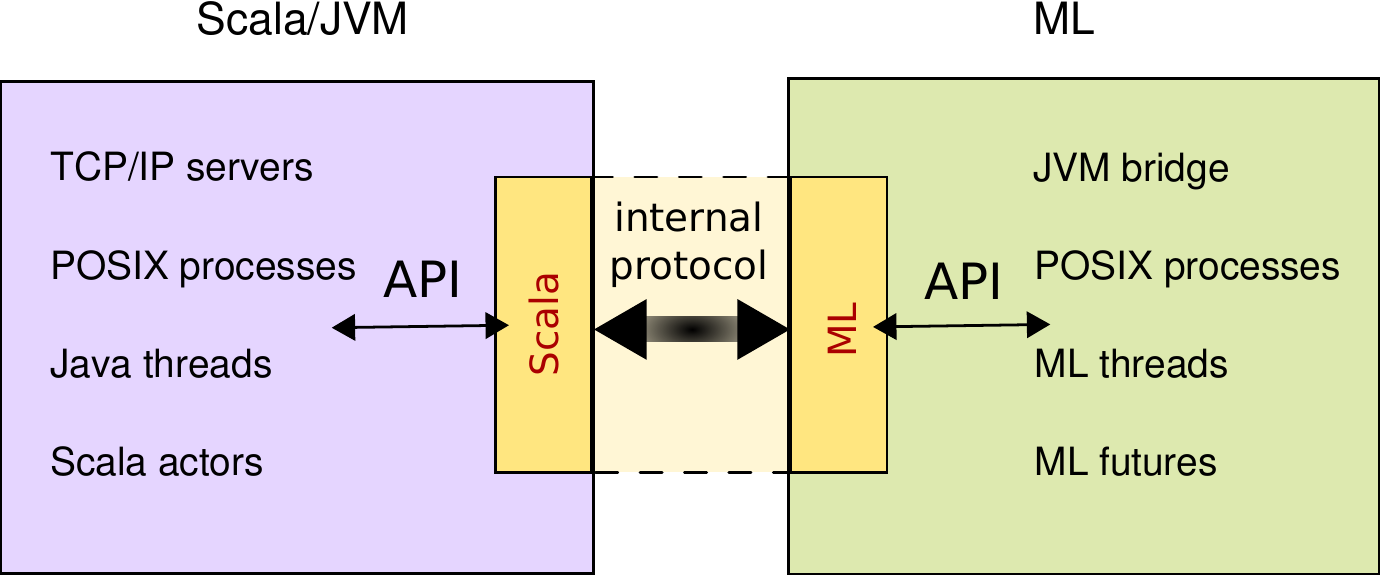}
  \end{center}

  The Isabelle/Scala interaction model is \emph{asynchronous}: the
  front-end can send batches of document updates to the prover, which
  will be processed in pipelined and parallel mode.  Results are
  accumulated in monotonic fashion in a persistent document model,
  which consists of collections of XML trees that are incrementally
  extended for each document version, and can be visualized on demand
  according to physical GUI events.  Document updates are pure
  mathematical operations that work on a family of versions with
  immutable content.

  Our mixed ML/Scala architecture allows to re-use decades of research
  into core prover technology, while integrating with more mainstream
  front-ends.  A typical example is the concrete Prover IDE as
  implemented in Isabelle/jEdit \cite{Wenzel:2010}.  Here the existing
  editor framework jEdit\footnote{\url{http://www.jedit.org}}, which
  is based on standard Java/Swing together with some simple plugin
  architecture, is connected to the semantic document content provided
  by the Isabelle process in the background.  Thus the user gets an
  impression of continuous proof checking, with IDE-style
  visualization of error messages, proof states, results etc.
  Additional semantic information is attached to the source text, and
  can be displayed as tooltips, popups, hyperlinks etc.\ within the
  editor.

  Isabelle/jEdit can be understood as reference application for other
  projects.  At the same time it is already usable as IDE for Isabelle
  theory development, which includes embedded Isabelle/ML for add-ons
  tools.  The following screenshot illustrates the immediate
  Isabelle/Isar/ML IDE aspect.

  \begin{center}
  \medskip
  \includegraphics[scale=0.4]{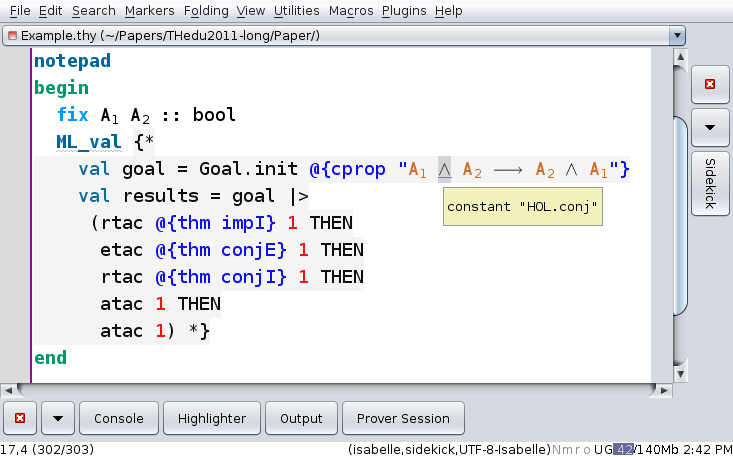}
  \medskip
  \end{center}

  The ML snippet within the logical notepad above refers to some
  logical entity embedded into the program text (``certified
  proposition'' \verb,@,\verb,{cprop},).  The term structure is
  annotated by semantic content from the prover using common
  document-oriented markup infrastructure of PIDE.  This enables a
  hyperlink to the formal entity of Isabelle/HOL that corresponds to
  the notation for conjunction ``\isa{{\isaliteral{5C3C616E643E}{\isasymand}}}''.  This connection of
  formal content with surface syntax works reliably despite somewhat
  complex nesting of several user-defined formal languages within the
  Isabelle/Isar framework.
  
  The underlying Poly/ML
  compiler\footnote{\url{http://www.polyml.org}} participates in the
  same document model of Isabelle/Scala, to provide immediate
  interactive feedback via semantic highlighting for keywords,
  sub-expressions and tooltips for inferred types.  This has required
  modest changes of the compiler by its author David Matthews.  Other
  implementations of ``domain-specific formal languages'' may
  participate in a similar manner, after minimal reforms of such classic
  command-line oriented tools.

  Physical rendering within the editor uses whatever happens to be
  available on the Java/Swing platform: mathematical symbols from the
  Unicode repertoire, font-styles with affine transformations (for
  sub/superscripts), colors with transparency (alpha-channel) etc.
  This approximates poor-man's mathematical type-setting within the
  bounds of the plain-text editor engine of jEdit.%
\end{isamarkuptext}%
\isamarkuptrue%
\isamarkupsection{Document-oriented prover interaction%
}
\isamarkuptrue%
\begin{isamarkuptext}%
Conceptually, the main abstraction of the PIDE framework is an
  interactive document-model that provides a \emph{timeless~/
  stateless} view on the results of the parallel and asynchronous
  proof engine in the background.  The main aspects of
  document-oriented prover interaction are as follows.

  \begin{itemize}

  \item Physical editor events stemming from conventional GUI
  components are turned into mathematical entities with explicit
  version information (algebra of document changes).  This continuous
  stream of changes is processed incrementally, with some intermediate
  pipelining and preprocessing.  The detachment from physical time and
  space enables sophisticated scheduling, parallel processing etc.\ on
  the prover side.

  \item Corresponding prover results are streamed towards the
  front-end.  Versions are identified explicitly, and content of
  common sub-structures of related document versions is shared to some
  extent.  The Isabelle/Scala document model absorbs all markup that
  is produced by the prover in an incremental manner.  The editor can
  query that information at any time, and interpret its XML content in
  terms of conventional GUI metaphors.

  \item Asynchronous streaming back and forth between the two
  processes naturally leads to time shifts and delays in the
  propagation of information.  The document API includes a notion of
  \emph{outdated} content and some approximation of future results by
  reverting text edits on old versions.  The implementation of the
  two-sided protocol engine works generally in a lock-free manner.
  Data access is \emph{not synchronized}, but certain values
  \emph{converge} to the intended meaning, by accumulating
  approximative information over time in a monotonic fashion.

  \item Document-content is persistent by default.  This means that
  conceptually, old version are never destroyed.  The implementation
  eventually disposes unreachable parts of the history to reclaim
  memory (garbage collection).  In particular, current Isabelle/jEdit
  merely maintains a relatively narrow window on versions that are
  being pipelined between the two processes.  So far there is no
  connection to the editor history: \emph{undo} or \emph{redo}
  operations on the text will reconstruct content by replaying
  relevant parts within the prover.

  \end{itemize}

  The subsequent screenshot illustrates the mathematical idea of
  \emph{document snapshot} within the physical editor.  The content is
  shown in various projections (via interpretation functions), using
  standard GUI metaphors as well as direct program access in the Scala
  console that is part of Isabelle/jEdit.

  \begin{center}
  \includegraphics[scale=0.3]{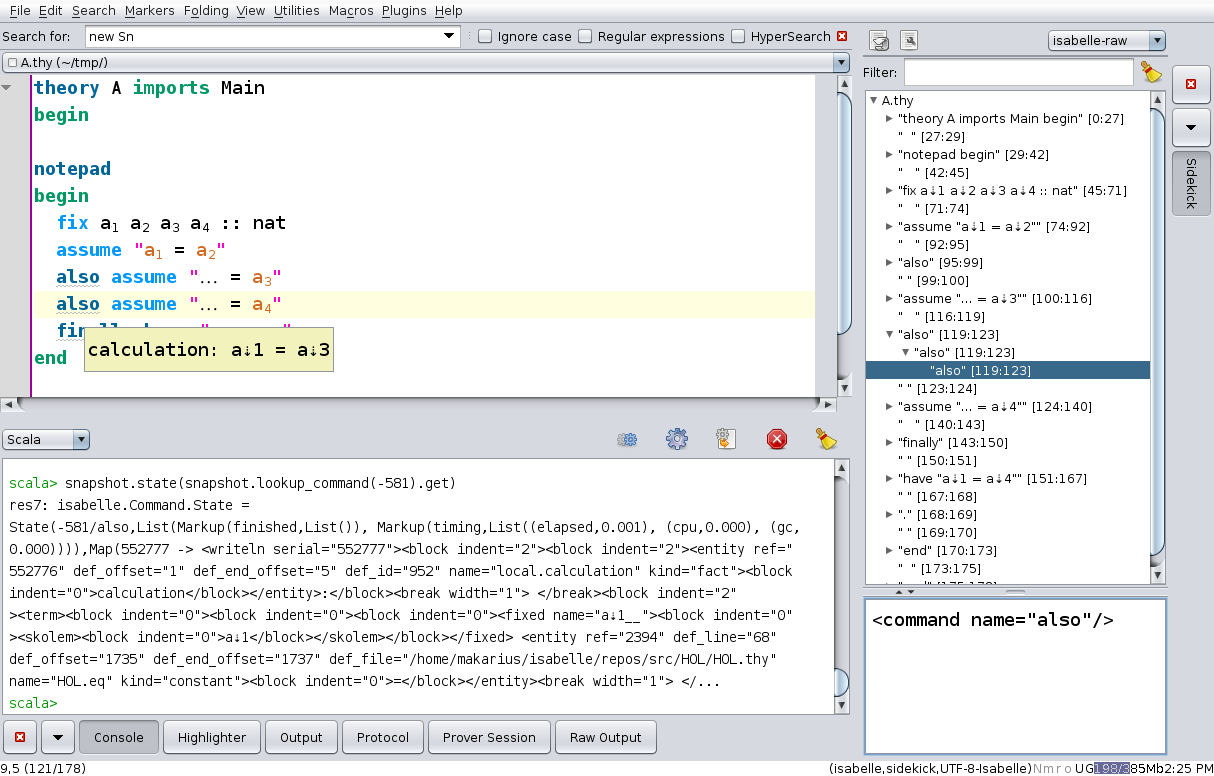}
  \end{center}

  The assembly of dockable panels in Isabelle/jEdit is managed within
  the same JVM process.  Conventional multithreading achieves smooth
  operation, while other editor tasks continue.\footnote{Contrast this
  to the single-threaded LISP interpreter of Emacs.}  The prover
  process is fully detached: neither the editor nor the prover are
  ever forced to wait for the other side.

  \medskip User tools implemented within the ML world of the prover
  may participate in the document-oriented approach by minimal means.
  The general programming model is a continuous extension of the TTY
  model: ML code runs within a certain transaction context, so regular
  text messages can be associated with a certain execution state of
  particular document versions.  The result will be associated with
  the document content in the proper position, thanks to
  mathematically captured space and time.

  Furthermore, the traditional text operations for prover output can
  be augmented to support full XML markup, by using our refined YXML
  transfer syntax \cite[\S2.3]{Wenzel:2011:CICM}, which allows text
  and markup to be composed orthogonally as byte sequence.  Due to
  augmented standard display operations of the prover, user code will
  automatically benefit from rich markup information that can be
  post-processed on the front-end side.

  \medskip The following example illustrates how the prover is able to
  speak XML natively, despite the appearance of plain-old
  text-oriented application code.%
\end{isamarkuptext}%
\isamarkuptrue%
\isadelimML
\endisadelimML
\isatagML
\isacommand{ML}\isamarkupfalse%
\ {\isaliteral{7B2A}{\isacharverbatimopen}}\isanewline
\ \ warning\ {\isaliteral{28}{\isacharparenleft}}{\isaliteral{22}{\isachardoublequote}}Term{\isaliteral{3A}{\isacharcolon}}\ {\isaliteral{22}{\isachardoublequote}}\ {\isaliteral{5E}{\isacharcircum}}\ Syntax{\isaliteral{2E}{\isachardot}}string{\isaliteral{5F}{\isacharunderscore}}of{\isaliteral{5F}{\isacharunderscore}}term\ ctxt\ t\ {\isaliteral{5E}{\isacharcircum}}\ Position{\isaliteral{2E}{\isachardot}}str{\isaliteral{5F}{\isacharunderscore}}of\ pos{\isaliteral{29}{\isacharparenright}}\isanewline
{\isaliteral{2A7D}{\isacharverbatimclose}}%
\endisatagML
{\isafoldML}%
\isadelimML
\endisadelimML
\begin{isamarkuptext}%
\medskip\noindent This prints term \verb,t, which is \isa{x\ {\isaliteral{2B}{\isacharplus}}\ y} containing some undeclared variables that Isabelle highlights
  to warn the user.  There is also some reference to the logical
  constant behind the ``\isa{{\isaliteral{2B}{\isacharplus}}}'' notation.  The Isabelle/Scala
  document-model receives the following XML tree:%
\end{isamarkuptext}%
\isamarkuptrue%
\begin{isamarkuptext}%
{\footnotesize
\begin{verbatim}
<warning serial="553408" offset="1" end_offset="3" id="124">Term:
<term><block indent="0"><block indent="0"><hilite><block
indent="0"><free><block indent="0">x</block></free></block></hilite>
<entity ref="20935" def_line="134" def_offset="3517"
def_end_offset="3522" def_file="~~/src/HOL/Groups.thy"
name="Groups.plus_class.plus" kind="constant"><block
indent="0">+</block></entity><break width="1"> </break><hilite><block
indent="0"><free><block
indent="0">y</block></free></block></hilite></block></block></term><position
offset="69" end_offset="72" id="-114"/></warning>
\end{verbatim}}%
\end{isamarkuptext}%
\isamarkuptrue%
\begin{isamarkuptext}%
Here we also see that the traditional pretty-printing of terms
  according to Oppen \cite{Oppen:1980} is turned into symbolic markup.
  Physical formatting is done later within some GUI container, based
  on precise window size and font-metrics --- no more fixed-width
  ASCII text on the TTY.

  \medskip The above idea of augmenting plain-text output by
  document-oriented markup depends on some minimal requirements of
  existing ML user code: standard Isabelle/ML functions to format term
  structures and to output text messages need to be used exclusively.
  Direct writing to stdio channels of the operating system is not
  appropriate.\footnote{In early attempts with PGIP the protocol
  machine would crash, but PIDE uses a private byte channel between
  the two processes and leaves stdin/stdout/stderr open for
  exceptional situations.  Thus raw stdout steps outside the document
  model, producing an accidental side-effect as physical process I/O.
  The front-end might capture that and display it on some system
  console for diagnostic purposes, or just ignore it.}

  Isabelle/ML also provides some explicit markup operations, such that
  user code may indicate additional semantic aspects in the text.
  Furthermore the classic canon of \verb|writeln|, \verb|warning| and
  \verb|error| messages is extended by \verb|Output.status| and \verb|Output.report|, which augment document content without immediate
  display to the user.  The latter is used for semantic
  syntax-highlighting of the sources, for example.%
\end{isamarkuptext}%
\isamarkuptrue%
\isamarkupsection{Further Application Scenarios%
}
\isamarkuptrue%
\begin{isamarkuptext}%
The ongoing development of the Isabelle/Scala layer and the
  Isabelle/jEdit application demonstrate that it is feasible to build
  combined mathematical assistants with LCF-style formal basis.  We
  can anticipate the following further application scenarios, which
  are particularly relevant for educational tool development.

  \begin{enumerate}

  \item Document preparation with rich semantic content, based on the
  logical context of a given theory library.  The idea is to render
  the markup information that the prover provides for formal source
  text into XHTML/CSS, or similar technologies.  This would allow to
  browse annotated text within a JVM-based HTML browser.  A typical
  application could be an \emph{interactive text-book}, where
  high-quality rendering is combined with some querying or exploration
  of formal prover content.

  \item Client-side mathematical editors with specific support for
  typical patterns like calculational reasoning and induction.  Such
  mathematical worksheets should also address proper mathematical
  notation, instead of the direct source presentation of the existing
  Prover IDE.  This requires some means for high-quality typesetting
  of mathematical formulae on the JVM, which are unfortunately not
  easily available in open-source, though.

  \item Rich-client IDE components as extensions to the Prover IDE.
  The design of jEdit makes it easy to assemble JVM/Swing components
  as plugins.  Thus existing tools like Geogebra that happen to work
  with the same platform can be easily integrated, at least at the GUI
  level.  The actual logical connection to some theory of geometry is
  a different issue, e.g.\ see \cite{Pham-Bertot:2010} for recent work
  with Coq.

  \item Server-side applications using JVM-based web frameworks.
  Java/Swing GUI components for rich clients are adequate, but the
  true strength of the JVM platform are its server components.  We
  have only started to investigate the possibilities: JVM-based web
  servers like Apache Tomcat or Jetty are commonplace; sophisticated
  application frameworks like Lift\footnote{\url{http://liftweb.net}}
  or Play\footnote{\url{http://www.playframework.org/}} are built on
  top, and allow easy integration of Scala components in particular.

  \end{enumerate}

The web application scenario might turn out particularly interesting
for educational purposes, especially in combination with the aspects
of document preparation and mathematical editors.  For example,
Isabelle/Scala + Lift could be used as a basis for some mathematical
Wiki \cite{mathwiki:2010} that can be used by students immediately
within their usual browser, tablet computer, or smart-phone.  Rich
clients like our existing Prover IDE fit into this scenario as
high-end authoring tools to produce formal libraries in the first
place.%
\end{isamarkuptext}%
\isamarkuptrue%
\isamarkupsection{Conclusion%
}
\isamarkuptrue%
\begin{isamarkuptext}%
The main contributions of the Isabelle/PIDE platform are some
  answers to old issues of prover interaction and integration.  The
  combination of ML/Scala successfully bridges the technological (and
  cultural) gap of higher-order functional and object-oriented
  programming.  This results in improved accessibility of
  fully-foundational LCF-style provers.  The particular benefits for
  educational scenarios are a corollary from that, among other
  possibilities not covered here (e.g.\ tools for formal methods).%
\end{isamarkuptext}%
\isamarkuptrue%
\isamarkupsubsection{Related work%
}
\isamarkuptrue%
\begin{isamarkuptext}%
The problem of prover integration has been tackled many times
  in the past.  Often the idea is to define each program as
  independent ``component'' with explicit communication protocols, say
  between provers, computer algebra systems, front-ends, etc.  This
  approach has been followed in the PROSPER project
  \cite{PROSPER:2003} and PGIP \cite{Aspinall-et-al:2007}, for example.

  We argue that cutting components at process boundary leads to rather
  complex interfaces defined by the protocols, which need to be
  standardized, implemented and maintained.  Some of this was
  attempted for PGIP and Isabelle in 2004--2006, but with limited
  success.  The postulated properties of the protocol and the reality
  of the prover did not fit together.  Moreover, the full potential of
  the prover was inhibited by the sequential / synchronous interaction
  model of PGIP, which was inherited from classic Proof~General.

  In Isabelle/PIDE these limitations have been overcome by cutting the
  main conceptual building blocks differently: a simple Scala API for
  declarative document editing, which is implemented by a complex
  protocol that bridges Scala/ML internally.  This scheme allows to
  follow the ongoing evolution of prover technology.  For example,
  when first versions of PGIP were drafted in 2004, the multicore
  revolution of 2005/2006 was not yet anticipated.  Parallel Isabelle
  is routinely available since 2008/2009
  \cite{Wenzel:2009,Matthews-Wenzel:2010}, but prover interaction
  protocols were lagging behind for a long time.

  \medskip Prover interaction and prover interfaces in the
  LCF-tradition usually means TTY commands, but this has been
  challenged occasionally in the past, e.g.\ in CtCoq/Pcoq
  \cite{Bertot-Thery:1998} or PLAT\(\Omega\) \cite{Wagner-et-al:2006}.
  Matita \cite{Asperti-et-al:2007} is probably the best-known proof
  assistant that was designed with some IDE support (based on
  OCaml/GTk) and advanced presentation formats (MathML) from the
  start, although its interaction model imitates classic Proof~General
  / Emacs \cite{Aspinall:2000} again.

  ProofWeb \cite{Kaliszyk:2006} transfers the Proof~General model to
  the web, using standard AJAX web technology: a custom-made webserver
  in OCaml is wrapped around the prover process on the server, the
  client uses JavaScript on Firefox.  This web application has been
  successfully applied for logic courses for university students
  \cite{hendriks-adn10}.%
\end{isamarkuptext}%
\isamarkuptrue%
\isamarkupsubsection{Future work%
}
\isamarkuptrue%
\begin{isamarkuptext}%
We expect that the current state of the Isabelle/Scala
  integration layer in Isabelle2011-1 is only the start of renewed
  interest in advanced prover integration scenarios.  An important
  trajectory for the existing document-model is scalability of the
  built-in editing history towards distributed version control with
  multiple users.

  This means, the current two-sided communication would eventually
  cover multiple end-points that work in a distributed manner
  world-wide.  Continuous formal checking could be backed by some
  server farm (or ``cloud''), managing many prover processes, each of
  them again running threads parallel.

  Further infrastructure will be required to organize collaborative
  users, which is especially interesting for education where teachers
  and students collaborate naturally.%
\end{isamarkuptext}%
\isamarkuptrue%
\isadelimtheory
\endisadelimtheory
\isatagtheory
\isacommand{end}\isamarkupfalse%
\endisatagtheory
{\isafoldtheory}%
\isadelimtheory
\endisadelimtheory
\isanewline
\end{isabellebody}%


\bibliographystyle{eptcs}
\bibliography{root}

\end{document}